\documentclass{article}
\usepackage{amsfonts}
\usepackage{amsmath}
\usepackage{cite}

\input{tcilatex}

\begin{document}

\title{New Interpretation for Laser Light Scattering Technique}
\author{Yong Sun{\thanks{%
Email: ysun2002h@yahoo.com.cn}}}
\maketitle

\begin{abstract}
The new method proposed in this work not only measures the particle size
distribution and the average molar mass accurately using the static light
scattering (SLS) technique when the Rayleigh-Gans-Debye approximation is
valid for dilute poly-disperse homogenous spherical particles in dispersion,
but also enables us to have insight into the theoretical analysis of the
dimensionless shape parameter $\rho $. With the method, a new size, static
radii $R_{s}$, can be measured. Based on the new static particle size
information, detailed investigation of the normalized time auto-correlation
function of the scattered light intensity $g^{\left( 2\right) }\left( \tau
\right) $ reveals that there exist three different particle sizes: the
static radius, hydrodynamic radius and apparent hydrodynamic radius that is
the hydrodynamic radius obtained using the cumulants method. With a simple
assumption that the hydrodynamic radius $R_{h}$ is in proportion to the
static radius $R_{s}$, the expected values of $g^{\left( 2\right) }\left(
\tau \right) $ calculated based on the static and commercial particle size
information are consistent with the experimental data. With the assistance
of simulated data, the apparent hydrodynamic radius is discussed. The
results show that the apparent hydrodynamic radius is different from the
mean hydrodynamic radius and is determined by the optical, hydrodynamic
characteristics and size distribution of particles and scattering vector.
The analysis also reveals that $\rho $ is determined by not only the
structure of particles but also the relationship between the optical and
hydrodynamic characteristics of particles even for mono-disperse model.
\end{abstract}

\section{INTRODUCTION}

For colloidal dispersion systems, light scattering is a widely used
technique to measure the sizes of particles. In dynamic light scattering
(DLS) technique, the standard method of cumulants\cite{re1,re2,re3,re4} has
been used to measure the hydrodynamic radius, or more strictly apparent
hydrodynamic radius $R_{h,app}$ \cite{re5} of particles from the normalized
time auto-correlation function of the scattered light intensity $g^{\left(
2\right) }\left( \tau \right) $ with the assistance of the Einstein-Stokes
relation, where $\tau $ is the delay time. $g^{\left( 2\right) }\left( \tau
\right) $ is considered to be determined by the hydrodynamic radius and the
scattering vector $q$ for dilute homogenous spherical particles\cite{re2}.
The treatment of the static light scattering (SLS) data is simplified to the
Zimm plot, Berry plot or Guinier plot etc. to get the root mean-square
radius of gyration $\left\langle R_{g}^{2}\right\rangle ^{1/2}$ and the
molar mass of particles provided that the particle sizes are small\cite%
{re4,re6}. For a long time, the measurements of the dimensionless shape
parameter $\rho =\left\langle R_{g}^{2}\right\rangle ^{1/2}/R_{h,app}$ \cite%
{re6,re7,re8,re9,re10} have been extensively used to infer the structures of
particles. In this judgement, it has an assumption that the particle sizes
measured from SLS and DLS are the same.

For large particles, DLS technique, where it loses the accuracy of size
measurements, is endeavored to use at different scattering vectors in order
to obtain the effective diffusion coefficient\cite{re11} or the apparent
hydrodynamic radius\cite{re5} to detect small poly-dispersities. The
standard DLS techniques are not suited to the accurate determination for the
poly-dispersities (standard deviation/mean size) less than about 10\%. For
dilute poly-disperse homogeneous spherical particles, Pusey and van Megen%
\cite{re11} proposed a method to detect small poly-dispersities when the
Rayleigh-Gans-Debye (RGD) approximation is valid for the particles with the
mean radius larger than 170 nm. In their treatment, the effective diffusion
coefficient is obtained from the initial slope of the logarithm of the
correlation function with respect to the scattering vector. By their
definition, the effective diffusion coefficient is an intensity-weighted
average diffusion coefficient. Both their theoretical and experimental
results show that the angular dependence of the effective diffusion
coefficient is a sensitive function of the particle size and distribution.

SLS technique has been reported to measure the particle size distribution by
some groups. Strawbridge and Hallett\cite{re12} studied the theoretical
scattered intensity of coated spheres with vertically polarized incident
light. The scattered intensity at the geometrical or linear trial radii
between $r_{\min }$ and $r_{\max }$ was used to fit the SLS data.
Schnablegger and Glatter\cite{re13} assumed that the particle size
distribution can be described as a series of cubic B-splines and used the
simulated and measured data to demonstrate the computation procedure.

In this work, a SLS treatment is reported for the size information of the
dilute poly-disperse homogeneous spherical particles in dispersion. The
number distribution of particle sizes is assumed to be Gaussian and the
effects of the scattering vector and the different intensity weights of
different particle sizes on the scattered light intensity are considered.
With the assistance of a non-linear least squares fitting program, the mean
static radius $\left\langle R_{s}\right\rangle $ and the standard deviation $%
\sigma $ are measured accurately. Given the absolute magnitude of the
scattered intensity and some parameters related to the instrument and
samples, the average molar mass can also be measured accurately. Based on
the static particle size information, detailed investigation of the
normalized time auto-correlation function of the scattered light intensity $%
g^{\left( 2\right) }\left( \tau \right) $ reveals that there exist three
different particle sizes: a static radius is measured from the optical
characteristics, a hydrodynamic radius is obtained from the hydrodynamic
features and an apparent hydrodynamic radius that is the hydrodynamic radius
obtained using the cumulants method is determined by the optical and
hydrodynamic characteristics of particles. With a simple assumption that the
hydrodynamic radius $R_{h}$ is in proportion to the static radius $R_{s}$,
the expected values of $g^{\left( 2\right) }\left( \tau \right) $ calculated
based on the static and commercial particle size information are consistent
with the experimental data. With the assistance of simulated data, $%
R_{h,app} $ is discussed. The results show that $R_{h,app}$ is different
from the mean hydrodynamic radius and is a composite size determined by the
optical, hydrodynamic characteristics and size distribution of particles and
scattering vector. The results also reveal that the theoretical value of $%
\rho $ is determined by not only the structure of particles, but also the
relationship between $R_{s}$ and $R_{h,app}$ even for mono-disperse model.

\section{THEORY}

For simplicity, poly-disperse homogeneous spherical particles are considered
and the RGD approximation is assumed to be valid. The average scattered
light intensity of a dilute non-interacting poly-disperse system in unit
volume can be obtained for vertically polarized light 
\begin{equation}
\frac{I_{s}}{I_{inc}}=\frac{4\pi ^{2}\sin ^{2}\theta _{1}n_{s}^{2}\left( 
\frac{dn}{dc}\right) _{c=0}^{2}c}{\lambda ^{4}r^{2}}\frac{4\pi \rho }{3}%
\frac{\int_{0}^{\infty }R_{s}^{6}P\left( q,R_{s}\right) G\left( R_{s}\right)
dR_{s}}{\int_{0}^{\infty }R_{s}^{3}G\left( R_{s}\right) dR_{s}},
\label{mainfit}
\end{equation}%
where $\theta _{1}$ is the angle between the polarization of the incident
electric field and the propagation direction of the scattered field, $c$ is
the mass concentration of particles, $r$ is the distance between the
scattering particle and the point of the intensity measurement, $\rho $ is
the density of the particles, $I_{inc}$ is the incident light intensity, $%
I_{s}$ is the intensity of the scattered light that reaches the detector, $%
R_{s}$ is the static radius of a particle, $\ q=\frac{4\pi }{\lambda }%
n_{s}\sin \frac{\theta }{2}$ is the scattering vector, $\lambda $ is the
wavelength of the incident light in vacuo, $n_{s}$\ is the solvent
refractive index, $\theta $ is the scattering angle, $P\left( q,R_{s}\right) 
$ is the form factor of homogeneous spherical particles

\begin{equation}
P\left( q,R_{s}\right) =\frac{9}{q^{6}R_{s}^{6}}\left( \sin \left(
qR_{s}\right) -qR_{s}\cos \left( qR_{s}\right) \right) ^{2}  \label{P(qr)}
\end{equation}
and $G\left( R_{s}\right) $ is the number distribution of particle sizes. In
this work, the number distribution is chosen as a Gaussian distribution

\begin{equation}
G\left( R_{s};\left\langle R_{s}\right\rangle ,\sigma \right) =\frac{1}{
\sigma \sqrt{2\pi }}\exp \left( -\frac{1}{2}\left( \frac{R_{s}-\left\langle
R_{s}\right\rangle }{\sigma }\right) ^{2}\right) ,
\end{equation}
where $\left\langle R_{s}\right\rangle $ is the mean static radius and $%
\sigma $ is the standard deviation.

\noindent The number average molar mass is defined as

\begin{equation}
\left\langle M\right\rangle =\frac{4\pi \rho }{3}N_A\int_{0}^{\infty
}R_{s}^{3}G\left( R_{s}\right) dR_{s},
\end{equation}
where $N_A$ represents Avogadro's number.

\noindent Comparing with the Zimm plot analysis, the mean square radius of
gyration $\left\langle R_{g}^{2}\right\rangle _{Zimm}$ for a poly-disperse
system is 
\begin{equation}
\left\langle R_{g}^{2}\right\rangle _{Zimm}=\frac{3\int_{0}^{\infty
}R_{s}^{8}G\left( R_{s}\right) dR_{s}}{5\int_{0}^{\infty }R_{s}^{6}G\left(
R_{s}\right) dR_{s}}.  \label{RG}
\end{equation}

\noindent If the reflected light is considered, the average scattered light
intensity in unit volume is written as 
\begin{equation}
\frac{I_{s}}{I_{inc}}=a\frac{4\pi \rho }{3}\frac{\int_{0}^{\infty
}R_{s}^{6}P\left( q,R_{s}\right) G\left( R_{s}\right)
dR_{s}+b\int_{0}^{\infty }R_{s}^{6}P\left( q^{\prime },R_{s}\right) G\left(
R_{s}\right) dR_{s}}{\int_{0}^{\infty }R_{s}^{3}G\left( R_{s}\right) dR_{s}}
\label{mainre}
\end{equation}
where

\begin{equation}
a=\frac{4\pi ^{2}\sin ^{2}\theta _{1}n_{s}^{2}\left( \frac{dn}{dc}\right)
_{c=0}^{2}c}{\lambda ^{4}r^{2}}
\end{equation}
and 
\begin{equation}
q^{\prime }=\frac{4\pi }{\lambda }n_{s}\sin \frac{\pi -\theta }{2}
\end{equation}
is the scattering vector of the reflected light. $b$ is a constant
determined by the shape of sample cell, the refractive indices of the
solvent and the sample cell and the geometry of instruments.

For dilute poly-disperse homogeneous spherical particles, the normalized
time auto-correlation function of the electric field of the scattered light $%
g^{\left(1\right)}(\tau )$ can be obtained

\begin{equation}
g^{\left( 1\right) }\left( \tau \right) =\frac{\int R_{s}^{6} P\left(
q,R_{s}\right)G\left( R_{s}\right) \exp \left( -q^{2}D\tau \right) dR_{s}}{%
\int R_{s}^{6}P\left( q,R_{s}\right) G\left( R_{s}\right) dR_{s}},
\label{Grhrs}
\end{equation}
where $D$ is the diffusion coefficient.

\noindent From the Einstein-Stokes relation

\begin{equation}
D=\frac{k_{B}T}{6\pi \eta _{0}R_{h}},
\end{equation}
where $\eta _{0}$, $k_{B}$ and $T$ are the viscosity of the solvent,
Boltzmann's constant and absolute temperature respectively, the hydrodynamic
radius $R_{h}$ can be obtained.

\noindent For simplicity, the relationship between the static and
hydrodynamic radii is assumed to be 
\begin{equation}
R_{h}=kR_{s},  \label{RsRh}
\end{equation}
where $k$ is a constant. From the Siegert relation between $g^{\left(
2\right) }\left( \tau \right) $ and $g^{\left( 1\right) }\left( \tau \right)$
\cite{re14}

\begin{equation}
g^{\left( 2\right) }\left( \tau \right) =1+\beta \left( g^{\left( 1\right)
}\right) ^{2},  \label{G1G2}
\end{equation}
the function between SLS and DLS is built and the values of $g^{\left(
2\right) }\left( \tau \right) $ can be expected based on the particle size
information obtained using the SLS technique and the values of $R_{h,app}$
also can be calculated.

\section{EXPERIMENT}

The SLS and DLS data were measured using the instrument built by ALV-Laser
Vertriebsgesellschaft m.b.H (Langen, Germany). It utilizes an ALV-5000
Multiple Tau Digital Correlator and a JDS Uniphase 1145P He-Ne laser to
provide a 23 mW vertically polarized laser at wavelength of 632.8 nm.

In this work, two kinds of samples were used. One is PNIPAM submicron
spheres. $N$-isopropylacrylamide (NIPAM, monomer) from Acros Organics was
recrystallized from hexane/acetone solution. Potassium persulfate (KPS,
initiator) and $N,N^{\prime }$-methylenebi-sacrylamide (BIS, cross-linker)
from Aldrich were used as received. Fresh de-ionized water from a Milli-Q
Plus water purification system (Millipore, Bedford, with a 0.2 $\mu m$
filter) was used throughout the experiments. The synthesis of gel particles
was described elsewhere\cite{re15,re16} and the recipes of the batches used
in this work are listed in Table \ref{table1}. The four samples were named
according to the molar ratios $n_{B}/n_{N}$ of $N,N^{\prime }$%
-methyle-nebisacrylamide over $N$-isopropylacrylamide.

\begin{table}[th]
\begin{center}
\begin{tabular}{|c|c|c|c|c|c|}
\hline
Sample & T ($^{\text{o}}$C) & $t$ (hrs) & $W_{N}+W_{B}$ (g) & $KPS$ (mg) & $%
n_{B}/n_{N}$ \\ \hline
PNIPAM-0 & $70\pm 1$ & $4.0$ & $1.00$ & $40$ & $0$ \\ \hline
PNIPAM-1 & $70\pm 1$ & $4.0$ & $1.00$ & $40$ & $1.0\%$ \\ \hline
PNIPAM-2 & $70\pm 1$ & $4.0$ & $1.00$ & $40$ & $2.0\%$ \\ \hline
PNIPAM-5 & $70\pm 1$ & $4.0$ & $1.00$ & $40$ & $5.0\%$ \\ \hline
\end{tabular}
\makeatletter
\par
\makeatother
\end{center}
\caption{Synthesis conditions for PNIPAM particles.}
\label{table1}
\end{table}

The four PNIPAM samples were centrifuged at 14,500 RPM followed by
decantation of the supernatants and re-dispersion in fresh de-ionized water
and process was repeated four times to remove free ions and any possible
linear chains. Then the samples were diluted for light scattering to weight
factors of $5.9\times 10^{-5}$, $8.56\times 10^{-6}$, $9.99\times 10^{-6}$
and $8.38\times 10^{-6}$ for PNIPAM-0, PNIPAM-1, PNIPAM-2 and PNIPAM-5
respectively. 0.45 $\mu m$ filters (Millipore, Bedford) were used to clarify
the samples PNIPAM-1, PNIPAM-2 and PNIPAM-5 before light scattering
measurements. The other kind of samples is two standard polystyrene latex
samples from Interfacial Dynamics Corporation (Portland, Oregon). One
polystyrene sample is the sulfate polystyrene latex with a normalized mean
radius of 33.5 nm (Latex-1) and the other is the surfactant-free sulfate
polystyrene latex of 55 nm (Latex-2), as shown in Table 2. Latex-1 and
Latex-2 were diluted for light scattering to weight factors of $1.02\times
10^{-5}$ and $1.58\times 10^{-5}$ respectively.

\section{DATA ANALYSIS}

In this section, the particle size information included in SLS and DLS and
the relationship between SLS and DLS are investigated

\subsection{Standard polystyrene latex samples}

The particle size information was provided by the supplier as obtained using
Transmission Electron Microscopy (TEM) technique. Because of the small
particle sizes and the large difference between the refractive indices of
the polystyrene latex (1.591 at wavelength 590 nm and 20 $^{\text{o}}$C) and
the water (1.332), i.e., the \textquotedblleft phase
shift\textquotedblright\ $\frac{4\pi }{\lambda }R|m-1|$ \cite{re4,re17} are
0.13 and 0.21 for Latex-1 and Latex-2 respectively, which do not exactly
satisfy the rough criterion for validity of the RGD approximation\cite{re4},
the mono-disperse model $G\left( R_{s}\right) =\delta \left(
R_{s}-\left\langle R_{s}\right\rangle \right) $ was used to measure the
approximate values of $\left\langle R_{s}\right\rangle $ for the two
polystyrene latex samples. The values of the mean radii and standard
deviations of the two samples shown in Table 2 were input into Eq. \ref%
{mainfit} to get $I_{s}/I_{inc}$ respectively. In order to compare with the
experimental data, the calculated value was set to be equal to that of the
experimental data at $q=0.0189$ nm$^{-1}$ for Latex-1. The results are shown
in Fig. 1a. In order to compare the expected root mean-square radii of
gyration $\left\langle R_{g}^{2}\right\rangle _{cal}^{1/2}$ with
experimental values $\left\langle R_{g}^{2}\right\rangle _{Zimm}^{1/2}$, the
values of the mean radii and standard deviations were input into Eq. \ref{RG}
to calculate $\left\langle R_{g}^{2}\right\rangle _{cal}^{1/2}$
respectively. Meanwhile $\left\langle R_{g}^{2}\right\rangle _{Zimm}^{1/2}$
was measured using the Zimm plot. Figure 1b shows the fit results of
Latex-1: $Kc/R_{vv}=1.29\times 10^{-8}+3.11\times 10^{-6}q^{2}$. The DLS
data of the two polystyrene latex samples were measured under the same
conditions as the SLS data respectively and $R_{h,app}$ was obtained using
the first cumulant method. For the two polystyrene latex samples, the values
of $R_{h,app}$ at a scattering vector of 0.00905 nm$^{-1}$ were chosen as
the results measured using the DLS technique since the values of $R_{h,app}$
almost do not depend on the scattering vector. All results are listed in
Table \ref{table2}. The results show that the difference between the static
and apparent hydrodynamic radii is large, the value measured using the SLS
technique is consistent with that measured using TEM and the expected value $%
\left\langle R_{g}^{2}\right\rangle _{cal}^{1/2}$ calculated using the
commercial size information is consistent with that measured using the Zimm
plot analysis.

\begin{table}[th]
\begin{center}
\begin{tabular}{|c|c|c|c|c|c|}
\hline
$\left\langle R\right\rangle _{comm}$ (nm) & $\sigma _{comm}$ (nm) & $%
\left\langle R_{g}^{2}\right\rangle _{Zimm}^{1/2}$ (nm) & $\left\langle
R_{g}^{2}\right\rangle _{cal}^{1/2}$ (nm) & $\left\langle R_{s}\right\rangle 
$ (nm) & $R_{h,app}$ (nm) \\ \hline
33.5(Latex-1) & 2.5 & 26.7 & 26.9 & 33.3$\pm $0.2 & 37.27$\pm $0.09 \\ \hline
55(Latex-2) & 2.5 & 46.8 & 43.2 & 56.77$\pm $0.04 & 64.5$\pm $0.6 \\ \hline
\end{tabular}%
\end{center}
\caption{The commercial size information, values of $\left\langle
R_{g}^{2}\right\rangle _{Zimm}^{1/2}$, $\left\langle R_{g}^{2}\right\rangle
_{cal}^{1/2}$, $\left\langle R_{s}\right\rangle $ and $R_{h,app}$ at a
scattering vector of 0.00905 nm$^{-1}$.}
\label{table2}
\end{table}

If the constant $k$ in Eq. \ref{RsRh} for Latex-1 is assumed to be 1.1 and
the size information provided by the supplier is assumed to be consistent
with that measured using the SLS technique, all the experimental data and
expected values of $g^{\left( 2\right) }\left( \tau \right) $ at a
temperature of 298.45 K and the scattering angles 30$^{\text{o}}$, 60$^{%
\text{o}}$, 90$^{\text{o}}$, 120$^{\text{o}}$ and 150$^{\text{o}}$ are shown
in Fig. 2. The expected values of $g^{\left( 2\right) }\left( \tau \right) $
are consistent with the experimental data.

If the expected values of $g^{\left( 2\right) }\left( \tau \right) $ were
calculated using Bargeron's equation\cite{re2}, all the experimental data
and expected values of $g^{\left( 2\right) }\left( \tau \right) $ for
Latex-1 at a temperature of 298.45 K and the scattering angles 30$^{\text{o}%
} $, 60$^{\text{o}}$, 90$^{\text{o}}$, 120$^{\text{o}}$ and 150$^{\text{o}}$
are shown in Fig. 3. The expected values of $g^{\left( 2\right) }\left( \tau
\right) $ have large differences with the experimental data.

\section{PNIPAM samples}

When Eq. \ref{mainfit} was used to fit the data of PNIPAM-1 measured at a
temperature of 302.33 K, it was found that the results of $\left\langle
R_{s}\right\rangle $ and $\sigma $ depend on the scattering vector range
being fit, as shown in Table \ref{table3}. If a small scattering vector
range is chosen, the parameters are not well-determined. As the scattering
vector range is increased, the uncertainties in the parameters decrease and $%
\left\langle R_{s}\right\rangle $ and $\sigma $ stabilize. If the fit range
continues to increase, the values of $\left\langle R_{s}\right\rangle $ and $%
\sigma $ begin to change and $\chi ^{2}$ grows. This is due to the deviation
between the experimental and theoretical scattered light intensity in the
vicinity of the scattered intensity minimum around $q=0.0177$ nm$^{-1}$,
where most of the scattered light is cancelled due to the light
interference. Many characteristics of particles could influence the
scattered light intensity in this region. For example, the number
distribution of particle sizes deviates from a Gaussian distribution, the
particle shapes deviate from a perfect sphere and the density of particles
deviates from homogeneity, etc. In order to avoid the effects of light
interference, the stable fit results $\left\langle R_{s}\right\rangle
=254.3\pm 0.1$ nm and $\sigma =21.5\pm 0.3$ nm obtained in the scattering
vector range from 0.00345 to 0.01517 nm$^{-1}$ are chosen as the size
information measured using the SLS technique. In order to examine the
influences of the fit ranges, fitting is also performed in an inverse way
where the largest value of $q$ in the fit range is fixed at 0.01517 nm$^{-1}$
while the smallest value of $q$ is varied. The fit results are also listed
in Table 3. The results show that $\left\langle R_{s}\right\rangle $ and $%
\sigma $ stabilize when the fit range is large enough. Figure 4 shows the
stable fit results and the residuals $\left( y_{i}-y_{fit}\right) /\sigma
_{i}$ in the scattering vector range from 0.00345 to 0.01517 nm$^{-1} $,
where $y_{i}$, $y_{fit}$ and $\sigma _{i}$ are the data, the fit value and
the uncertainty in the data at a given delay time $\tau _{i}$, respectively.

\begin{table}[tbp]
\begin{center}
\begin{tabular}{|c|c|c|c|}
\hline
$q$ ($10^{-3}$ nm$^{-1}$) & $\left\langle R_{s}\right\rangle $ (nm) & $%
\sigma $ (nm) & $\chi ^{2}$ \\ \hline
3.45 to 9.05 & 260.09$\pm $9.81 & 12.66$\pm $19.81 & 1.64 \\ \hline
3.45 to 11.18 & 260.30$\pm $1.49 & 12.30$\pm $3.37 & 1.65 \\ \hline
3.45 to 13.23 & 253.45$\pm $0.69 & 22.80$\pm $0.94 & 2.26 \\ \hline
3.45 to 14.21 & 254.10$\pm $0.15 & 21.94$\pm $0.36 & 2.03 \\ \hline
3.45 to 15.17 & 254.34$\pm $0.12 & 21.47$\pm $0.33 & 2.15 \\ \hline
3.45 to 17.00 & 255.40$\pm $0.10 & 17.32$\pm $0.22 & 11.02 \\ \hline
5.50 to 15.17 & 254.24$\pm $0.15 & 21.95$\pm $0.47 & 2.32 \\ \hline
7.95 to 15.17 & 254.32$\pm $0.16 & 21.56$\pm $0.57 & 2.38 \\ \hline
10.12 to 15.17 & 254.65$\pm $0.10 & 17.81$\pm $0.63 & 0.79 \\ \hline
12.21 to 15.17 & 254.84$\pm $0.16 & 19.33$\pm $0.87 & 0.42 \\ \hline
\end{tabular}
\makeatletter
\par
\makeatother
\end{center}
\caption{The fit results obtained using Eq. \protect\ref{mainfit} for
PNIPAM-1 at different scattering vector ranges and a temperature of 302.33
K. }
\label{table3}
\end{table}

When the reflected light was considered, Eq. \ref{mainre} was used to fit
all data in the full scattering vector range (0.00345 to 0.0255 nm$^{-1}$)
for the various factors of reflected light $b$. The fit results are listed
in Table \ref{table4}. The results show that the values of $\chi ^{2}$ are
much larger. The mean static radius $\left\langle R_{s}\right\rangle $ is
consistent with that measured using Eq. \ref{mainfit} in the scattering
vector range from 0.00345 to 0.01517 nm$^{-1}$ and the standard deviation
changes to smaller.

\begin{table}[tbp]
\begin{center}
\begin{tabular}{|c|c|c|c|}
\hline
$b$ & $\left\langle R_{s}\right\rangle$ (nm) & $\sigma$ (nm) & $\chi ^{2}$
\\ \hline
0.01 & 254.0$\pm $0.3 & 14.4$\pm $0.5 & 194.60 \\ \hline
0.011 & 254.0$\pm $0.3 & 14.6$\pm $0.5 & 168.20 \\ \hline
0.012 & 254.0$\pm $0.3 & 14.7$\pm $0.5 & 149.99 \\ \hline
0.013 & 254.0$\pm $0.2 & 14.8$\pm $0.4 & 139.82 \\ \hline
0.014 & 254.1$\pm $0.2 & 15.0$\pm $0.4 & 137.52 \\ \hline
0.015 & 254.1$\pm $0.2 & 15.1$\pm $0.4 & 142.96 \\ \hline
0.016 & 254.09$\pm $0.07 & 15.2$\pm $0.5 & 155.97 \\ \hline
0.017 & 254.1$\pm $0.3 & 15.4$\pm $0.5 & 176.40 \\ \hline
0.018 & 254.1$\pm $0.3 & 15.5$\pm $0.5 & 204.08 \\ \hline
\end{tabular}
\makeatletter
\par
\makeatother
\end{center}
\caption{The fit results for PNIPAM-1 obtained from Eq. \protect\ref{mainre}
using the various values of $b$.}
\label{table4}
\end{table}

As discussed above, light interference in the vicinity of the scattered
intensity minimum would influence the fit results. In order to eliminate the
effects of light interference, the experimental data in the vicinity of the
scattered intensity minimum were neglected. Eq. \ref{mainre} was thus used
to fit the experimental data in the full scattering vector range again. The
fit values are listed in Table \ref{table5}. The mean static radius and
standard deviation are consistent with the stable fit results obtained using
Eq. \ref{mainfit} in the scattering vector range from 0.00345 to 0.01517 nm$%
^{-1}$.

\begin{table}[tbp]
\begin{center}
\begin{tabular}{|c|c|c|c|}
\hline
$b$ & $\left\langle R_{s}\right\rangle$ (nm) & $\sigma$ (nm) & $\chi ^{2}$
\\ \hline
0.013 & 251.3$\pm $0.6 & 22.17$\pm $0.05 & 79.80 \\ \hline
0.014 & 251.1$\pm $0.6 & 23.3$\pm $0.9 & 58.29 \\ \hline
0.015 & 250.9$\pm $0.6 & 24.4$\pm $0.8 & 44.50 \\ \hline
0.016 & 250.7$\pm $0.5 & 25.4$\pm $0.7 & 37.02 \\ \hline
0.017 & 250.5$\pm $0.6 & 26.4$\pm $0.7 & 36.01 \\ \hline
0.018 & 250.3$\pm $0.6 & 27.24$\pm $0.8 & 41.59 \\ \hline
\end{tabular}
\makeatletter
\par
\makeatother
\end{center}
\caption{The fit results for PNIPAM-1 obtained using Eq. \protect\ref{mainre}
and neglecting experimental data near the intensity minimum.}
\label{table5}
\end{table}

If the constant $k$ in Eq. \ref{RsRh} for the PNIPAM-1 is assumed to be
1.21, all the experimental data and expected values of $g^{\left( 2\right)
}\left( \tau \right) $ at the scattering angles 30$^{\text{o}}$, 50$^{\text{o%
}}$ and 70$^{\text{o}}$ are shown in Fig. 5. The expected values of $%
g^{\left( 2\right) }\left( \tau \right) $ are consistent with the
experimental data.

If the expected values of $g^{\left( 2\right) }\left( \tau \right) $ were
calculated using Bargeron's equation, all the experimental data and expected
values of $g^{\left( 2\right) }\left( \tau \right) $ for PNIPAM-1 at the
scattering angles 30$^{\text{o}}$, 50$^{\text{o}}$ and 70$^{\text{o}}$ are
shown in Fig. 6. The expected values of $g^{\left( 2\right) }\left( \tau
\right) $ have large differences with the experimental data.

Traditionally the particle size information is measured using the DLS
technique. The standard method is the cumulants or the inverse Laplace
transform. For the five experimental data of $g^{\left( 2\right) }\left(
\tau \right) $ measured under the same conditions as the SLS data, their
corresponding fit results using the first cumulant and first two cumulants
respectively for PNIPAM-1 at a temperature of 302.33 K and a scattering
angle of 30$^{\text{o}}$ are listed in Table \ref{table6}.

\begin{table}[tbp]
\begin{center}
\begin{tabular}{|c|c|c|c|c|c|}
\hline
& $\left\langle \Gamma \right\rangle _{first}$ & $\chi ^{2}$ & $\left\langle
\Gamma \right\rangle _{two}$ & $\mu _{2}$ & $\chi ^{2}$ \\ \hline
1 & 39.73 $\pm $0.07 & 0.07 & 39.9 $\pm $0.1 & 28.20$\pm $15.99 & 0.04 \\ 
\hline
2 & 39.49 $\pm $0.07 & 0.33 & 40.2 $\pm $0.1 & 90.10$\pm $17.11 & 0.04 \\ 
\hline
3 & 39.86 $\pm $0.07 & 0.11 & 40.2 $\pm $0.1 & 39.17$\pm $16.19 & 0.05 \\ 
\hline
4 & 39.70 $\pm $0.07 & 0.07 & 39.9 $\pm $0.1 & 20.93$\pm $15.92 & 0.06 \\ 
\hline
5 & 39.34 $\pm $0.07 & 0.53 & 40.2 $\pm $0.1 & 112.75$\pm $17.26 & 0.08 \\ 
\hline
\end{tabular}%
\end{center}
\caption{The fit results of $g^{\left( 2\right) }\left( \protect\tau \right) 
$ for PNIPAM-1 at a temperature of 302.33 K and a scattering angle of 30$%
^\mathrm o$.}
\label{table6}
\end{table}

From the fit results, the values of the mean decay rate $\left\langle \Gamma
\right\rangle $ show an independence on the measurements, but the results of 
$\mu _{2}$ have a strong dependence on them. The values of $\mu _{2}$ are
often negative. It's a contradiction with its definition. In order to avoid
the contradiction that the values of $\mu _{2}$ depend on the DLS
measurements, the values of $R_{h,app}$ are obtained directly using the
first cumulant. The result of $R_{h,app}$ at a scattering angle 30$^{\text{o}%
}$ is 322.$\pm $2. nm. The difference between $R_{h,app}$ and $\left\langle
R_{s}\right\rangle $ is large.

For the PNIPAM samples at high temperatures, the situation using Eq. \ref%
{mainfit} is the same as that of PNIPAM-1 at a temperature of 302.33 K. The
values of $\left\langle R_{s}\right\rangle $ and $\sigma $ depend on the
scattering vector range being fit. If a small scattering vector range is
chosen, the parameters are not well-determined. As the scattering vector
range is increased, the uncertainties in the parameters decrease and $%
\left\langle R_{s}\right\rangle $ and $\sigma $ stabilize. The stable fit
results $\left\langle R_s\right\rangle = 139.3 \pm 0.3$ nm and $\sigma =
12.4 \pm 0.6$ nm obtained in the scattering vector range from 0.00345 to
0.02555 nm$^{-1}$ for PNIPAM-5 at a temperature of 312.66 K are chosen as
the size information measured using the SLS technique. Figure 7 shows the
stable fit results and the residuals.

If the constant $k$ in Eq. \ref{RsRh} for PNIPAM-5 is assumed to be 1.1, all
the experimental data and expected values of $g^{\left( 2\right) }\left(
\tau \right) $ at the scattering angles 30$^{\text{o}}$, 50$^{\text{o}}$, 70$%
^{\text{o}}$ and 100$^{\text{o}}$ are shown in Fig. 8. The expected values
of $g^{\left( 2\right) }\left( \tau \right) $ are consistent with the
experimental data. The value of $R_{h,app}$ at a scattering angle 30$^{\text{%
o}}$ is 158.9$\pm $0.7 nm. The difference between $R_{h,app}$ and $%
\left\langle R_{s}\right\rangle $ also is large.

\section{RESULTS AND DISCUSSION}

Same conclusions are also obtained for all other samples investigated. The
fit results of $\left\langle R_{s}\right\rangle $ and $\sigma $ depend on
the scattering vector range being fit. If a small scattering vector range is
chosen, the parameters are not well-determined. As the scattering vector
range is increased, the uncertainties in the parameters decrease and $%
\left\langle R_{s}\right\rangle $ and $\sigma $ stabilize.

For the PNIPAM samples, the main reason for the difference between
experimental and theoretical scattered intensity in the vicinity of the
scattered intensity minimum seems to be that the number distribution of
particle sizes deviates from a Gaussian distribution. With the mean static
radius and standard deviation obtained using Eq. \ref{mainfit} in the $q$
range between 0.00345 and 0.01517 nm$^{-1}$ from the SLS data of PNIPAM-1
measured at a temperature of 302.33 K, three different ways of calculation
were performed to see which can give the best expectation of the
experimental data. In Fig. 9, the expected values of the scattered intensity
related to incident intensity were first calculated using Eq. \ref{mainfit}
in the full particle size distribution range between 1 and 800 nm. The
calculated curve (solid line) matches the experimental date points only when
q is smaller than 0.016 nm$^{-1}$. Then, a truncated Gaussian distribution
was used and the calculation was performed between the $\left\langle
R_{s}\right\rangle -1.3\sigma $ and $\left\langle R_{s}\right\rangle
+1.3\sigma $ using Eq. \ref{mainfit}. The calculated curve (dash line)
matches the experimental date points in a broad $q$ range including the
vicinity of the scattered intensity minimum and deviates only at $q \geq
0.021$ nm$^{-1}$ where the reflected light could be detected. Finally, the
integrated range did not change but the reflected light was considered and
Eq. \ref{mainre} was used to calculate the expected results assuming $%
b=0.014 $. The calculated curve (dot line) matches the experimental date in
all $q$ range investigated. The results show that the scattered intensity in
the vicinity of the scattered intensity minimum is very sensitive to the
particle size distribution and the influences of the reflected light only
lie at very large scattering vectors.

The difference between $\left\langle R_{s}\right\rangle $ and $R_{h,app}$ is
large, showing evidences that the different particle sizes for a particle
system can be obtained using the light scattering technique. From the
theoretical analysis of $\left\langle R_{g}^{2}\right\rangle _{Zimm}^{1/2}$,
the dimensionless parameter $\left\langle R_{g}^{2}\right\rangle
_{Zimm}^{1/2}/ \left\langle R_s \right\rangle $ is determined by the
structure of particles and particle size distribution. For mono-disperse
homogenous spherical particles, the theoretical value of $R_g / R_s $ is
0.775. Due to the effects of the particle size distribution, the value of $%
\left\langle R_{g}^{2}\right\rangle _{Zimm}^{1/2}/ \left\langle R_s
\right\rangle $ is larger than 0.775. In the analysis of $\rho$, due to the
fact that $R_{h,app}$ and $\left\langle R_{s}\right\rangle $ are different
quantities, the value of $\rho$ is also determined by the relationship
between them. For the two polystyrene latex samples Latex-1 and Latex-2, the
values of $\rho$ are 0.716 and 0.726, respectively.

From the analysis of $g^{\left( 2\right) }\left( \tau \right)$, the expected
values of $g^{\left( 2\right) }\left( \tau \right)$ calculated using Eqs \ref%
{Grhrs}, \ref{RsRh} and \ref{G1G2} based on the static and commercial
particle size information are consistent with the experimental data. The
results also reveal that different particle size information is included in $%
g^{\left( 2\right) }\left( \tau \right)$. In order to discuss this question
conveniently, the simulated data are used.

The simulated data were produced using the information: the mean static
radius $\left\langle R_{s}\right\rangle $, standard deviation $\sigma $,
temperature $T$, viscosity of the solvent $\eta _{0}$, scattering angle $%
\theta $, wavelength of laser light $\lambda $, refractive index of the
water $n_{s}$ and constant $k$ were set to 260 nm, 26 nm, 302.33K, 0.8132 mPa%
$\cdot $S, 30$^{\text{o}}$, 632.8 nm, 1.332 and 1.2, respectively. When the
data of $\left( g^{\left( 2\right) }\left( \tau \right) -1\right) /\beta $
were obtained, the 1\% statistical noises were added and the random errors
were set 3\%. Five simulated data were produced. The fit results for the
five simulated data using the first cumulant and first two cumulants are
listed in Table \ref{table7}.

\begin{table}[tbp]
\begin{center}
\begin{tabular}{|c|c|c|c|c|c|}
\hline
& $\left\langle \Gamma \right\rangle _{first}$ & $\chi ^{2}$ & $\left\langle
\Gamma \right\rangle _{two}$ & $\mu _{2}$ & $\chi ^{2}$ \\ \hline
1 & 39.634 $\pm $0.002 & 11.98 & 39.95 $\pm $0.01 & 22.0$\pm $0.6 & 7.75 \\ 
\hline
2 & 39.252 $\pm $0.004 & 4.56 & 39.49 $\pm $0.02 & 9.8$\pm $0.6 & 3.83 \\ 
\hline
3 & 39.173 $\pm $0.002 & 5.67 & 39.71 $\pm $0.03 & 20.6$\pm $1.2 & 4.66 \\ 
\hline
4 & 39.164 $\pm $0.004 & 25.40 & 39.13 $\pm $0.01 & -1.7$\pm $0.4 & 25.44 \\ 
\hline
5 & 39.297 $\pm $0.002 & 15.75 & 39.40 $\pm $0.01 & 4.9$\pm $0.5 & 15.55 \\ 
\hline
\end{tabular}%
\end{center}
\caption{The fit results of $g^{\left( 2\right) }\left( \protect\tau \right) 
$ for the simulated data with the standard deviation 26 nm.}
\label{table7}
\end{table}

From the fit results of the simulated data, the situation is the same as the
experimental data: the values of the mean decay rate $\left\langle \Gamma
\right\rangle $ show an independence on the different noises and errors and
the results of $\mu _{2}$ have a strong dependence on them. The values of $%
\mu _{2}$ can be negative. As discussed above, a truncated Gaussian
distribution can give better expectation for the SLS data of PNIPAM-1 at a
temperature of 302.33 K, so the five simulated data were produced again with
the truncated Gaussian distribution that the range of integral is 221 to 299
nm. The fit results for this five simulated data using the first cumulant
and first two cumulants respectively are shown in Table \ref{table8}. The
values of $\mu _{2}$ still have a strong dependence on the different noises
and errors and are often negative.

\begin{table}[tbp]
\begin{center}
\begin{tabular}{|c|c|c|c|c|c|}
\hline
& $\left\langle \Gamma \right\rangle _{first}$ & $\chi ^{2}$ & $\left\langle
\Gamma \right\rangle _{two}$ & $\mu _{2}$ & $\chi ^{2}$ \\ \hline
1 & 39.998 $\pm $0.001 & 10.96 & 39.867 $\pm $ 0.009 & -5.4$\pm $0.4 & 10.36
\\ \hline
2 & 39.914 $\pm $0.007 & 20.57 & 39.90 $\pm $ 0.03 & -0.95$\pm $1.24 & 20.63
\\ \hline
3 & 40.045 $\pm $0.002 & 5.30 & 40.35 $\pm $ 0.02 & 12.8$\pm $0.8 & 4.61 \\ 
\hline
4 & 39.963 $\pm $0.005 & 3.97 & 40.07 $\pm $ 0.02 & 4.3$\pm $0.6 & 3.84 \\ 
\hline
5 & 39.992 $\pm $0.003 & 9.00 & 40.241 $\pm $0.008 & 9.4$\pm $0.3 & 5.51 \\ 
\hline
\end{tabular}%
\end{center}
\caption{The fit results of $g^{\left( 2\right) }\left( \protect\tau \right) 
$ for the simulated data produced with a truncated distribution.}
\label{table8}
\end{table}

Comparing the fit results using the first cumulant with the values using the
first two cumulants for the experimental and simulated data respectively,
the values of the mean decay rate are consistent. In order to avoid the
contradiction that the values of $\mu _{2}$ depend on the DLS measurements, $%
R_{h,app}$ is measured using the first cumulant. Meanwhile, from the
theoretical analysis of cumulants, $R_{h,app}$ is obtained from averaging
the term $\exp \left( -q^{2}D\tau \right) $ in the static particle size
distribution $G\left( R_{s}\right) $ with the weight $R_{s}^{6}P\left(
q,R_{s}\right) $. In order to explore the effects of the particle size
distribution, the simulated data were produced as the simulated data above
with the same mean static radius 260 nm and the different standard
deviations 13, 39 and 52 nm respectively. The constant $k$ is still chosen
1.2. From this assumption, the mean hydrodynamic radius is 312 nm. The
apparent hydrodynamic radii obtained from $g^{\left( 2\right) }\left( \tau
\right) $ for different standard deviations are listed in Table \ref{table9}.

\begin{table}[tbp]
\begin{center}
\begin{tabular}{|c|c|}
\hline
$\sigma /\left\langle R_{s}\right\rangle $ & $R_{h,app}$ (nm) \\ \hline
5\% & 315.7$\pm $0.9 \\ \hline
10\% & 325.$\pm $2. \\ \hline
15\% & 339.4$\pm $0.9 \\ \hline
20\% & 356.$\pm $1. \\ \hline
\end{tabular}%
\end{center}
\caption{Values of $R_{h,app}$ for the simulated data produced using the
same mean static radius and different standard deviations.}
\label{table9}
\end{table}

The results reveal that the values of $R_{h,app}$ are influenced obviously
by the standard deviation. As shown in Eq. \ref{Grhrs}, the quantity $\exp
\left( -q^{2}D\tau \right) $ is determined by the hydrodynamic
characteristics while $R_{s}^{6}P\left( q,R_{s}\right) $ is determined by
the optical features of particles. As a result, $g^{\left( 2\right) }\left(
\tau \right) $ is determined by both the optical and hydrodynamic
characteristics of particles. When the cumulants method is used, $R_{h,app}$
obtained from $g^{\left( 2\right) }\left( \tau \right) $ is a composite size
determined by the optical, hydrodynamic characteristics and size
distribution of particles and scattering vector. If the simple size
information needs to be obtained from $g^{\left( 2\right) }\left( \tau
\right) $, the relationship between the optical and hydrodynamic quantities
must be considered. The accurate relationship between the static and
hydrodynamic radii can be explored further.

\section{CONCLUSION}

Eq. \ref{mainfit} provides a method to accurately measure particle size
distribution. Given the absolute magnitude of the scattered intensity and
some parameters related to the instrument and samples, the average molar
mass of large particles can also be measure accurately. Using the light
scattering technique, three different particle sizes can be measured. The
static radius is measured from the optical characteristics, the hydrodynamic
radius is obtained from the hydrodynamic features and the apparent
hydrodynamic radius is determined by the optical, hydrodynamic
characteristics and size distribution of particles and scattering vector.
With a simple assumption that the hydrodynamic radius $R_{h}$ is in
proportion to the static radius $R_{s}$, the expected values of $g^{\left(
2\right) }\left( \tau \right) $ calculated based on the static and
commercial particle size information are consistent with the experimental
data and the apparent hydrodynamic radius obtained using the cumulants
method is different from the mean hydrodynamic radius. The theoretical
values of dimensionless shape parameter $\rho $ is related to not only the
structure of particles, but also the relationship between the static radius
and the apparent hydrodynamic radius even for mono-disperse model. $%
g^{\left( 2\right) }\left( \tau \right) $ at a scattering vector contains
the optical and hydrodynamic information of particles. If the accurate
relationship between the optical and hydrodynamic quantities can be
understood, the static particle size information can also be measured
accurately from DLS.

Fig. 1. a). The experimental data and expected values of $I_{s}/I_{inc}$ and
b). The Zimm plot for Latex-1. In a, the circles show the experimental data
and the line represents the expected values of $I_{s}/I_{inc}$. In b, the
circles show the experimental data and the line shows a linear fit to the
plot of $Kc/R_{vv}$ as a function of $q^2$.

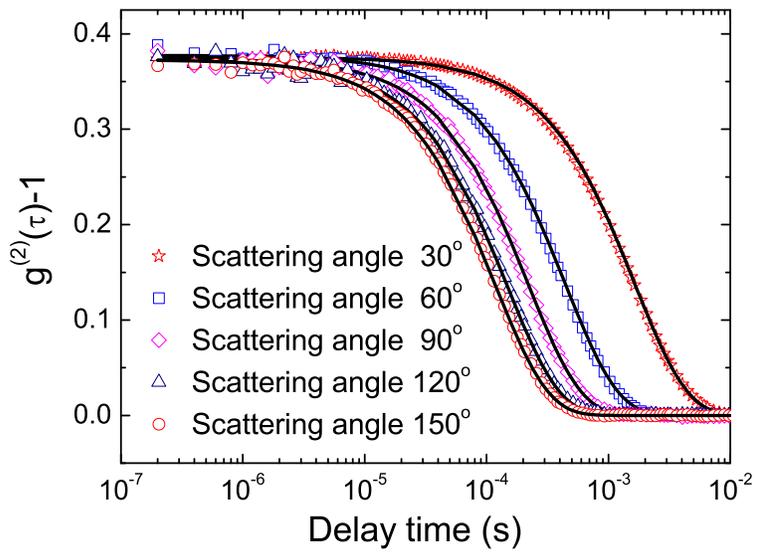
Fig. 2. The experimental data and expected values of $g^{\left( 2\right)
}\left( \tau \right) $ for Latex-1. The symbols show the experimental data
and the lines show the expected values calculated under the simple
assumption $R_{h}=1.1R_{s}$.

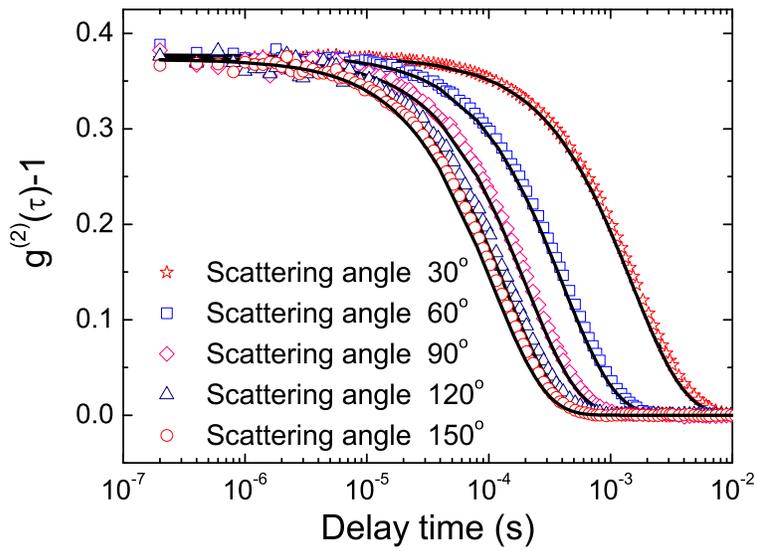
Fig. 3. The experimental data and expected values of $g^{\left( 2\right)
}\left( \tau \right) $ for Latex-1. The symbols show the experimental data
and the lines show the expected values calculated under the simple
assumption $R_{h}=R_{s}$.

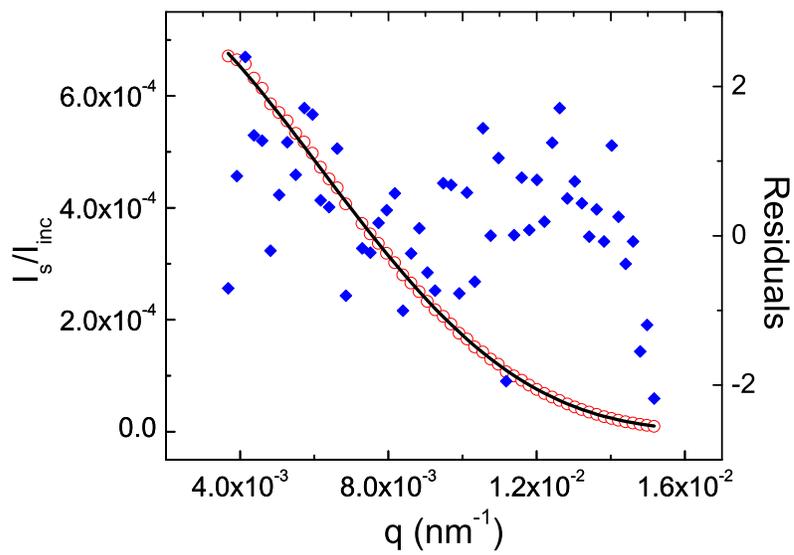
Fig. 4. The experimental data and stable fit results obtained using Eq. \ref%
{mainfit} for PNIPAM-1 at a temperature of 302.33 K. The circles show the
experimental data, the line shows the fit results and the diamonds show the
residuals.

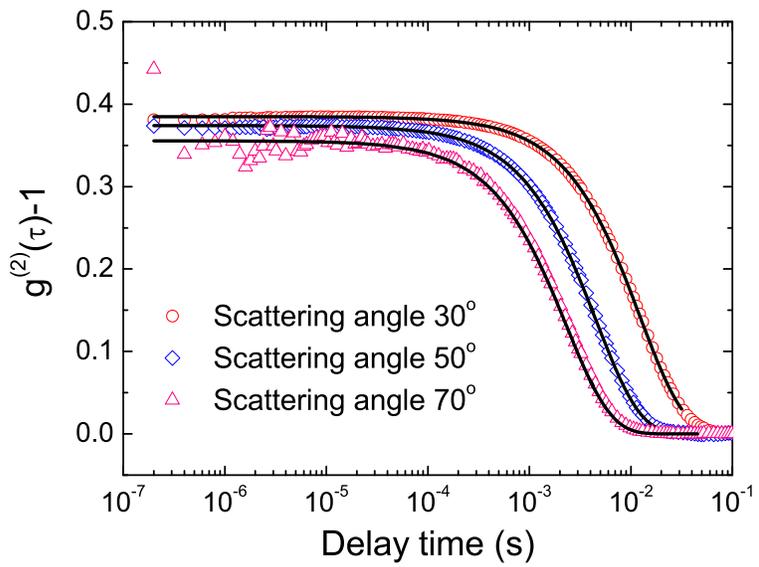
Fig. 5. The experimental data and expected values of $g^{\left( 2\right)
}\left( \tau \right)$ for PNIPAM-1 at a temperature of 302.33 K. The symbols
show the experimental data and the lines show the expected values calculated
under the simple assumption $R_{h}=1.21R_{s}$.

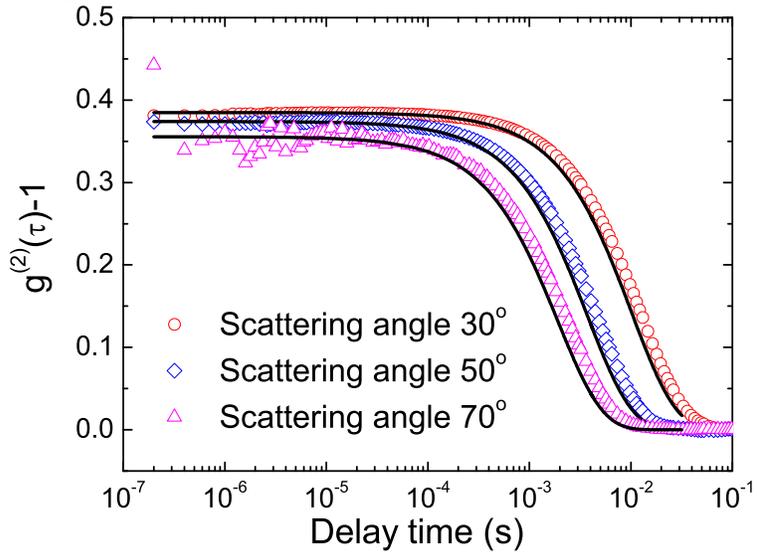
Fig. 6. The experimental data and expected values of $g^{\left( 2\right)
}\left( \tau \right)$ for PNIPAM-1 at a temperature of 302.33 K. The symbols
show the experimental data and the lines show the expected values calculated
under the simple assumption $R_{h}=R_{s}$.

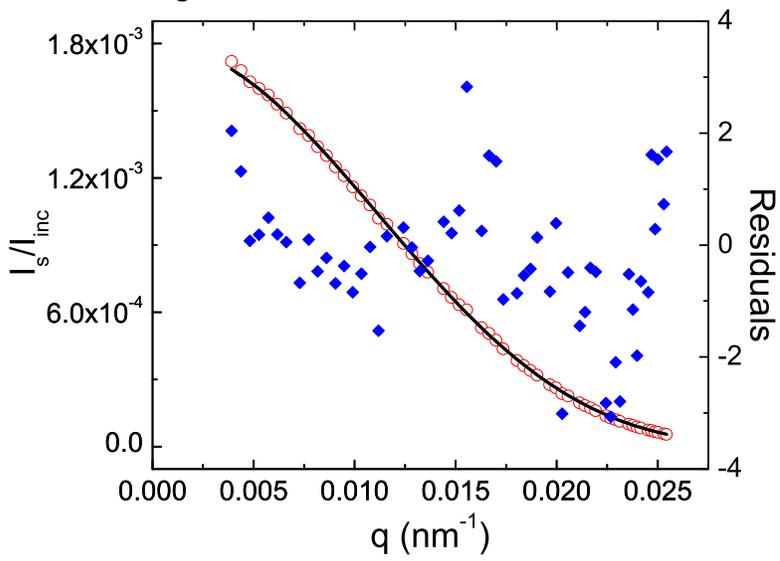
Fig. 7. The experimental data and stable fit results obtained using Eq. \ref%
{mainfit} for PNIPAM-5 at a temperature of 312.66 K. The circles show the
experimental data, the line shows the fit results and the diamonds show the
residuals.

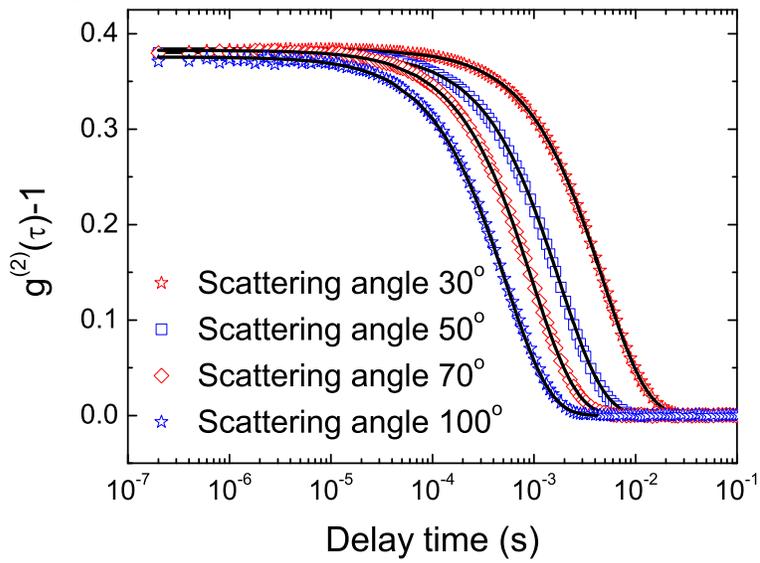
Fig. 8. The experimental data and expected values of $g^{\left( 2\right)
}\left( \tau \right)$ for PNIPAM-5 at a temperature of 312.66 K. The symbols
show the experimental data and the lines show the expected values calculated
under the simple assumption $R_h=1.1R_s$.

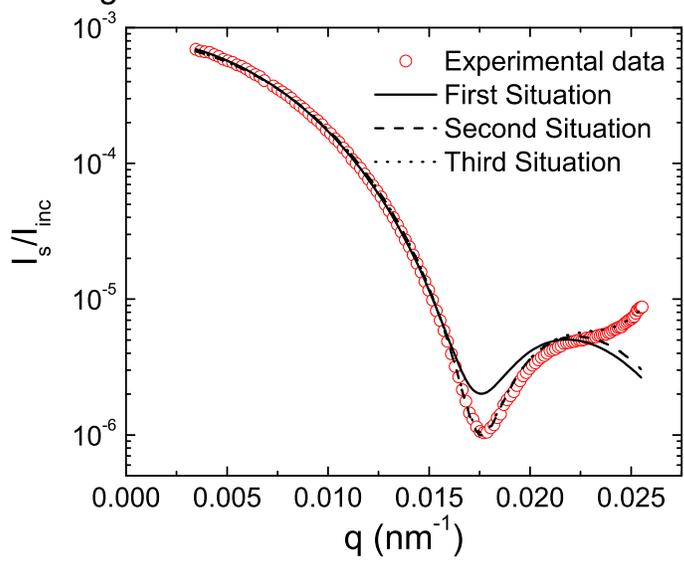
Fig. 9. The experimental data and expected values for PNIPAM-1 at a
temperature of 302.33 K. The circles show the experimental data, the solid
line shows the expected values calculated using Eq. \ref{mainfit} in the
full particle size distribution range, the dash line represents the expected
values calculated using Eq. \ref{mainfit} between about the $\left\langle
R_{s}\right\rangle -1.3\sigma $ and $\left\langle R_{s}\right\rangle
+1.3\sigma $ and the dot line shows the expected values calculated using Eq. %
\ref{mainre} in the same range as the second with $b$: $0.014$.

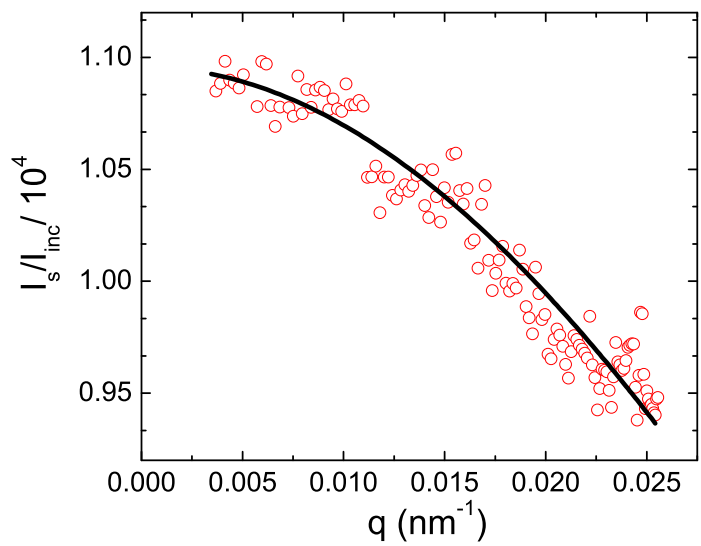
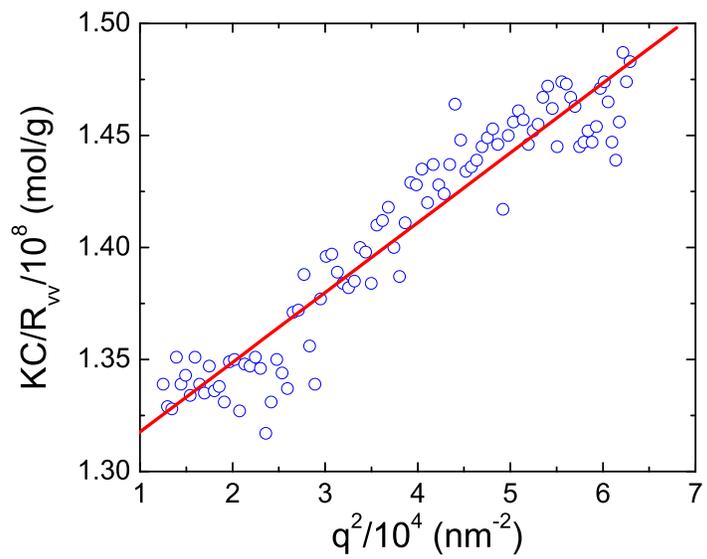
\end{document}